\documentclass[twocolumn,letterpaper,10pt]{article}
\usepackage{iasted}
\usepackage{times}
\usepackage[dvips]{graphicx}

\usepackage{subfigure}
\usepackage{makeidx}
\usepackage{hyperref}
\usepackage{algorithm}
\usepackage{algorithmic}

\usepackage{multirow}
\usepackage{amsmath}
\usepackage{amssymb}
\usepackage{xcolor}
\usepackage{epsfig}
\usepackage{graphics}
\begin{document}

\date{12/26/2012}

\title{Optimizing Synchronization Algorithm for Auto-parallelizing Compiler}

\author{Gang~Liao, Si-hui~Qin, Long-fei~Ma, Qi~Sun\\Department of Computer Science and Engineering, Sichuan University Jinjiang College, China, Pengshan, 620860.
\\greenhat1016@gmail.com, zhihuiqindoris@gmail.com, fly3601@gmail.com, SunQi3737@gamil.com}
\maketitle
\thispagestyle{empty}

\noindent
{\bf\normalsize ABSTRACT}\newline
{In this paper, we focus on the need for two approaches to optimize producer and consumer synchronization for auto-parallelizing compiler. Emphasis is placed on the construction of a criterion model by which the compiler reduce the number of synchronization operations needed to synchronize the dependence in a loop and perform optimization reduces the overhead of enforcing all dependence. In accordance with our study, we transform to modify and eliminate dependence on iteration space diagram (ISD), and carry out the problems of acyclic and cyclic dependence in detail. we eliminate partial dependence and optimize the synchronize instructions. Some didactic examples are included to illustrate the optimize  procedure.} \vspace{2ex}

\noindent
{\bf\normalsize KEY WORDS}\newline
{auto-parallelizing, compiler, synchronization, dependence analysis}

\section{Introduction}

During the past decade, the field of compiling for parallel architecture has exploded with  widespread commercial availability of multicore processors \cite{1}\cite{2}. Research has focused on several goals, the major concern being support for auto-parallelizing. The goal of auto-parallelizing is compiling an invariant and unannotated sequential program into a parallel program \cite{3}.

Although in recent years most attention has been given to support for languages with parallel annotations (i.e. OpenMP \cite{4} allow programmer to manually hint compiler about parallel regions.), the parallelization of legacy code still has a profound historical significance. The Parafrase system \cite{5} is the first automatic parallelize compiler based on dependence analysis, which was developed at the University of Illinois. The most ambitious for parafrase was to find out how to develop architecture to exploit the latent parallelism in off-the-shelf dusty deck programs \cite{6}.
By using producer/consumer synchronization (e.g. the Alliant F/X8 \cite{8} \cite{9} implemented synchronization instructions), this ordering can be forced on the program execution, allowing parallelism to be extracted from loops with dependence.

In this paper, We focus on the parallelization of legacy code and optimizing producer/consumer synchronization via two approaches in auto-parallelizing compiler. We proceed as follows. First, in section 2, we present the compiler fundamentals and the target architecture. In order to understand the latter section, we introduce some concepts of auto-parallelizing compiler so as to be acquainted with jargons. In additional, for clarity and brevity are served by directing the discussion towards a single architecture. In section 3, in order to understand how parallelism can be extracted from cyclic loops using producer/consumer synchronization, we must discuss how to extract parallelism when the dependence graph may be cyclic and loop freezing cannot be used to break the cycles. In section 4, we show how to reduce and  optimize the number of synchronization instructions used to synchronize a loop.

\section{The Compiler Fundamentals and Target Computer}
In order to relieve programmers from the tedious and error-prone manual parallelize process, the compiler need automatic convert sequential code into multi-threaded or vectorization code to utilize multiple processors simultaneously in a shared-memory multiprocessors machine.

\subsection{Automatic Parallelize Compiler Fundamentals}
\begin{figure}
\centering
\includegraphics[width=0.50\textwidth]{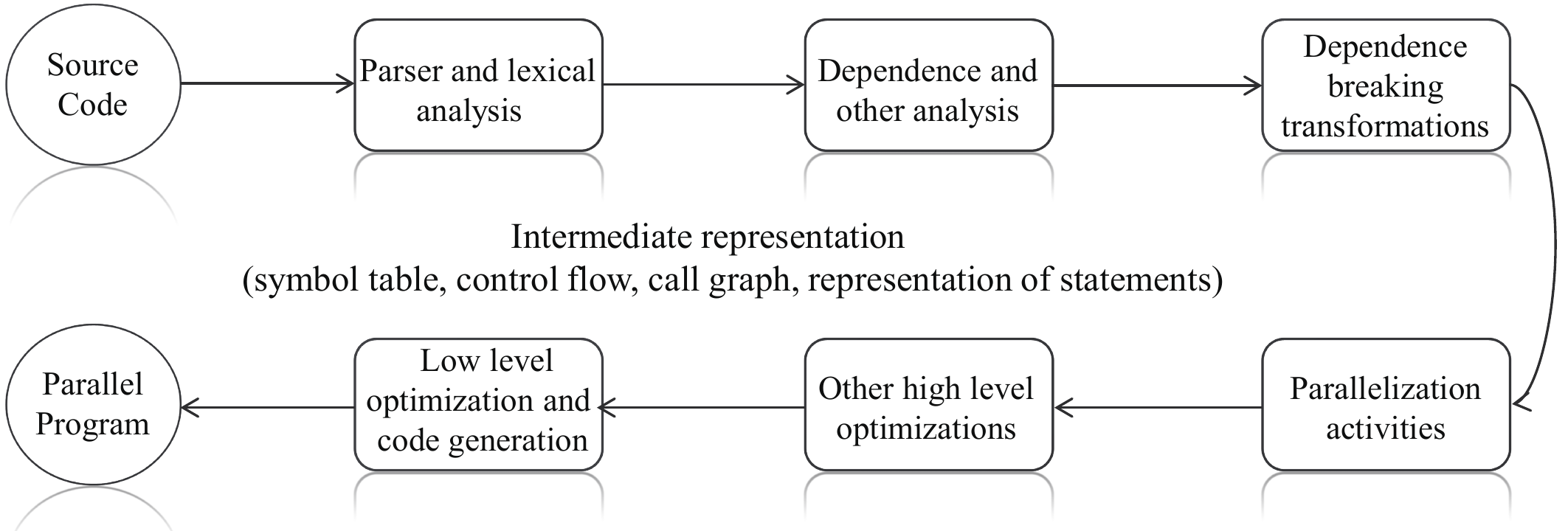}
\caption{High-level structure of a parallel compiler}
\label{fig1}
\end{figure}

The high level flow of a compiler is shown in Figure.\ref{fig1}. The actual phases of the compiler are shown as the centre, as well as inputs and intermediate files are shown as rounded boxes.

In fact, the source program may be a binary file, used in binary instruments and binary compilers \cite{11}. In general, a Java or Python source-to-byte code compiler would convert the binary file to the $bytecode$ file which contains analysis information for the $compilation$ $unit$ included, to the further dependence analysis on a compilation unit. A compilation unit is $lexically$ $analyzed$ and $parsed$ by the compiler. The lexical analysis and parsing are not studied in this paper.
A discussion of detailed techniques for compiler can be found in \cite{12} (e.g. regular expression, deterministic finite automata, non-deterministic finite automata). The result of the parser is an $intermediate$ $representation$ (IR), which is regarded as an $abstract$ $syntax$ $tree$ and a graphical representation of the parsed program. We will modify this slightly and represent programs as a $control$ $flow$ $graph$ (CFG).
In a control flow graph, each node $b_i \in B$ is $basic$ $block$. There are, in most presentations, two specially designated blocks: the entry block, through which control enters into the flow graph, and the exit block, through which all control flow leaves. Where an edge $b_i \rightarrow b_j$ means that $b_i$ may execute directly before $b_j$. In additional, A CFG are sometimes converted to $static$ $single$ $assignment$ (SSA) form \cite{13}.

Dependence analysis determines whether or not it is safe to reorder or parallel statements. In general,
control dependence ($S_1  \boldsymbol{\delta}^c  S_2$) is a situation in which a program's instruction executes if the previous instruction evaluates in a way that allows its execution. A data dependence ($S_1  \boldsymbol{\delta}^f  S_2$, $S_1  \boldsymbol{\delta}^a  S_2$, $S_1  \boldsymbol{\delta}^o  S_2$, $S_1  \boldsymbol{\delta}^i  S_2$ ) arises from two statements which access or modify the same resource \cite{7}. Loop dependence analysis is mostly done to find ways to do auto-parallelizing, which is the task of determining whether statements within a loop body form a dependence, with respect to array access and modification, induction, reduction and private variables, simplification of loop-independent code and management of conditional branches inside the loop body.

\subsection{Shared Memory Multiprocessors Machine}

\begin{figure}
\centering
\includegraphics[width=0.50\textwidth]{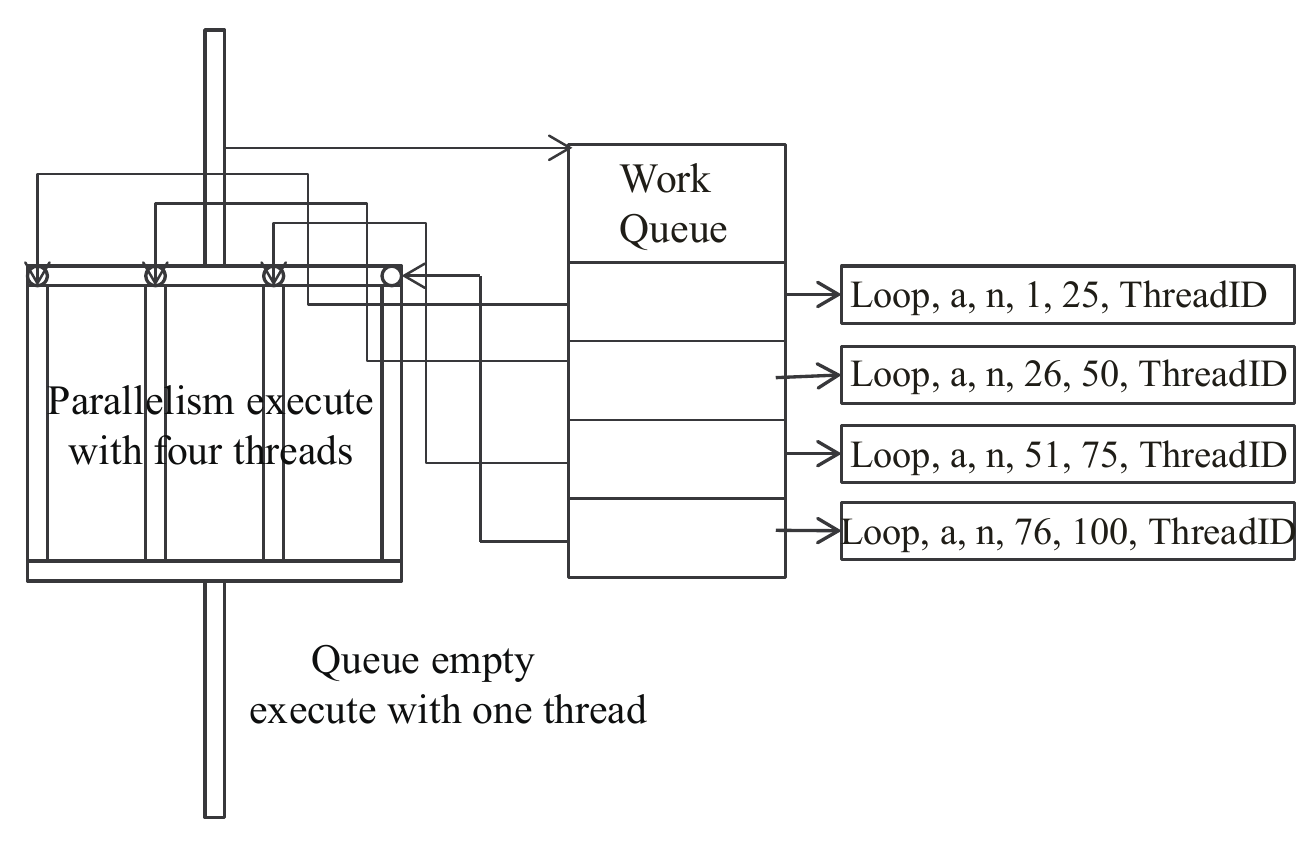}
\caption{Scheduling and execution of a shared memory program}
\label{fig2}
\end{figure}

In order to clarity and brevity, the target computer assumed throughout this paper is a shared memory multiprocessor. In these
systems, the processing elements can access any of the global memory modules through an interconnection network and code executes serially on each processor, and parallelism is realized by the simultaneous execution of different iterations of a loop on different processors. In the shared memory version of the program, each thread executes a subset of the iteration space of a parallel loop. The Cartesian space define slightly the boundary of the loop for the loop's iteration space.
In Figure.\ref{fig2} an example of scheduling and execution of a shared memory program is shown.
However, all large machines for high-performance numerical computing have a physically distributed memory architecture. The distributed memory machines consist of nodes connected to one another by using Ethernet or a variety proprietary interfaces.

Here, We presented a short, informal discussion of compiler fundamentals and shared memory multiprocessor machine. The interested reader will find a more complete discussion in \cite{12}\cite{14}. In the latter section, the details of producer/consumer synchronize optimizations would be discussed in this paper.

\section{Acyclic and Cyclic Dependence Analysis}
Most of the transformations in this paper are based on the concept of dependence between statements. In a sequential program, the statement instance $S_b^j$ is $flow$ $dependence$ on the statement instance $S_a^i$ ($S_a^i  \boldsymbol{\delta}^f S_b^j$) if $S_a^i$ assigns a value to a variable that may later be read by $S_b^j$. $S_b^j$ is $antidependence$ on $S_a^i$ ($S_a^i  \boldsymbol{\delta}^a  S_b^j$) if $S_a^i$ fetches from a variable that may be later written by $S_b^j$. $S_b^j$ is $output$ $dependence$ on $S_a^i$ ($S_a^i  \boldsymbol{\delta}^o  S_b^j$) if $S_a^i$ modifies a variable that may be later modified by $S_b^j$. $S_b^j$ is $control$ $dependence$ on $S_a^i$ ($S_a^i  \boldsymbol{\delta}^c  S_b^j$) if $S_a^i$ is control construct, and whether $S_b^j$ executes or not depends on the outcome of $S_a^i$. The more detailed discussion can be found in \cite{15} \cite{16}.

In order to parallel loops with acyclic and cyclic dependence graphs, Samuel P. Midkiff summarized
the following steps will be performed \cite{17}.
A dependence graph would be constructed for the loop nest;
Find strongly connected components (SCC) formed by cycles of dependence in the graph, contract the nodes in the SCC into a single large node; (Note: a directed graph is called components of strongly connected if there is a path from each vertex in the graph to every other vertex.)
Mark all nodes in the graph containing a single statement as parallel;
All inter-node dependence are lexically forward via topologically sort;
Group independent, unordered, nodes reading the same data and marked as parallel into new nodes to optimize data reuse;
Carry out loop fission to constitute a new loop for each node;
Mark as parallel all loops resulting from nodes whose statements are marked as parallel in the sorted graph;

These steps will be explained in detail by means of an example in the remainder of this section.

\subsection{Parallelizing Loops with Acyclic}

\begin{figure}
\centering
\includegraphics[width=0.50\textwidth]{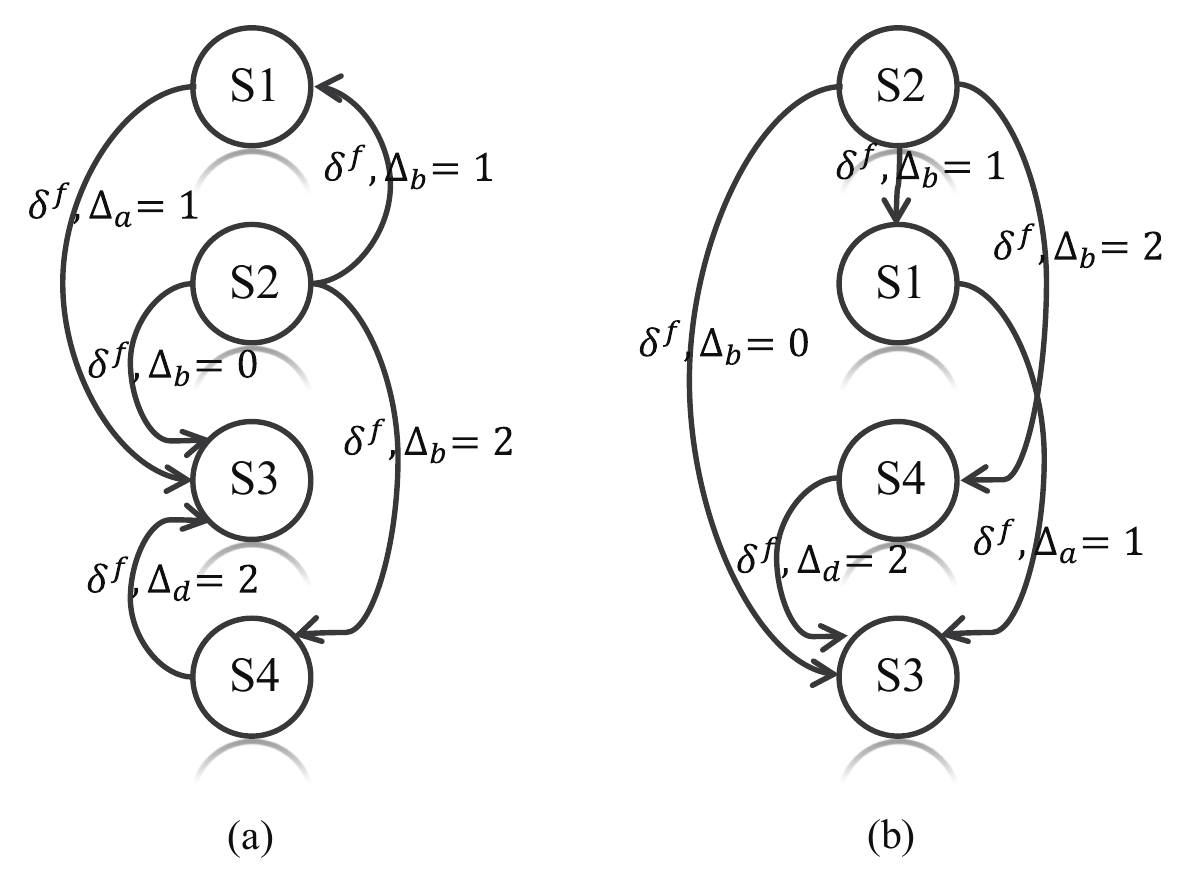}
\caption{(a) The dependence graph for the program. (b) The dependence graph after it has been topologically sorted.}
\label{fig3}
\end{figure}

A program with the dependence graph for a loop, as shown in Alg.\ref{alg1}. The acyclic dependence graph for the program is illustrated in Fig.\ref{fig3} (a). The $\Delta$ defines the dependence $distance$ (e.g. given a dependence $S_a^i \delta S_b^j$ between instances, $\Delta = j-i$). The node at the tail of a dependence arc is the dependence source($S_a$), and at the head of the arc is the dependence sink ($S_b$). In order to topologically sorting the dependence graph, all dependence must be $lexically$ $forward$ ($\Delta >= 0$. i.e. in branchless code the sink of the dependence is lexically forward of the source of the dependence). The canonical application of topological sorting is in scheduling a sequence of jobs or tasks based on their dependencies. A topological ordering is possible if and only if the graph has no directed cycles, that is, if it is a directed acyclic graph (DAG). Any DAG has at least one topological ordering, and the algorithm are known for constructing a topological ordering of any DAG in linear time. The more detailed algorithm can be found in \cite{18}.
\begin{algorithm}
\caption{A program with dependence.} \label{alg1}
\begin{algorithmic}
       \FOR{$i=1$; $i<n$; $i++$ }
          \item $S1: \space a[i] \gets b[i-1]+... ;$\\
           \item $S2: \space b[i] \gets c[i-1]+... ;$\\
           \item $S3: \space ... \gets a[i-1]+b[i]*d[i-2] ;$\\
          \item $S4: \space d[i] \gets b[i-2]-... ;$\\
         \ENDFOR
\end{algorithmic}
\end{algorithm}

Simultaneously, since code executes serially on a given processor, and therefore within an iteration of a loop, only dependence with a distance greater than zero ($\Delta >= 0$) need to be synchronized explicitly.

After the topological sorted, the dependence graph Fig.\ref{fig3} (a) is transformed to the Fig.\ref{fig3} (b). There are several possible ordering of the nodes resulting from a topological sort, that's one valid order. After that, the loop can be fully parallelized by breaking up the loop with dependence into multiple loops, none of which contain the source and sink of a loop carried (cross-iteration) dependence. The loop is transformed by reordering the statements to match the topological sort order, just like Alg.\ref{alg2}.

The program Alg.\ref{alg2} is a more efficient parallelization that can be performed by a different partitioning of statements
among loops that is still consistent with the ordering implied by the topological sort. In additional, the more efficient partitioning keeps statements that are not related by a loop-carried dependence together in the same loop.
 It called $loop$ $fission$ (also called $loop$ $distribution$ in the literature \cite{19}). Acyclic portions of the dependence graph may be sorted so that dependence are lexically forward, with a legal fission then being possible. In the program Alg.\ref{alg2}, S1 and S4 can remain in the same loop which is no loop-carried dependence. That is, the program with a statement ordering yielding slightly better locality, just like Alg.\ref{alg3}.

\begin{algorithm}
\caption{The program is transformed to reflect the order of the topologically sorted dependence graph} \label{alg2}
\begin{algorithmic}

      \FOR{$parallel$ $i=1$; $i<n$; $i++$}
         \item $S2: \space b[i] \gets c[i-1]+... ;$\\
       \ENDFOR

      \FOR{$parallel$ $i=1$; $i<n$; $i++$}
         \item $S1: \space a[i] \gets b[i-1]+... ;$\\
       \ENDFOR

      \FOR{$parallel$ $i=1$; $i<n$; $i++$}
         \item $S4: \space d[i] \gets b[i-2]-... ;$\\
       \ENDFOR

      \FOR{$parallel$ $i=1$; $i<n$; $i++$}
         \item $S3: \space ... \gets a[i-1]+b[i]*d[i-2] ;$\\
       \ENDFOR
\end{algorithmic}
\end{algorithm}

\begin{algorithm}
\caption{The program is transformed to reflect the order of the topologically sorted dependence graph and loop fission} \label{alg3}
\begin{algorithmic}
         \STATE $(invariant) ...$\\
        \FOR{$parallel$ $i=1$; $i<n$; $i++$ }
          \item $S1: \space a[i] \gets b[i-1]+... ;$\\
          \item $S4: \space d[i] \gets b[i-2]-... ;$\\
         \ENDFOR
         \STATE $(invariant) ...$\\
\end{algorithmic}
\end{algorithm}

\subsection{Parallelizing Loops with cyclic}
\begin{figure}
\centering
\includegraphics[width=0.40\textwidth]{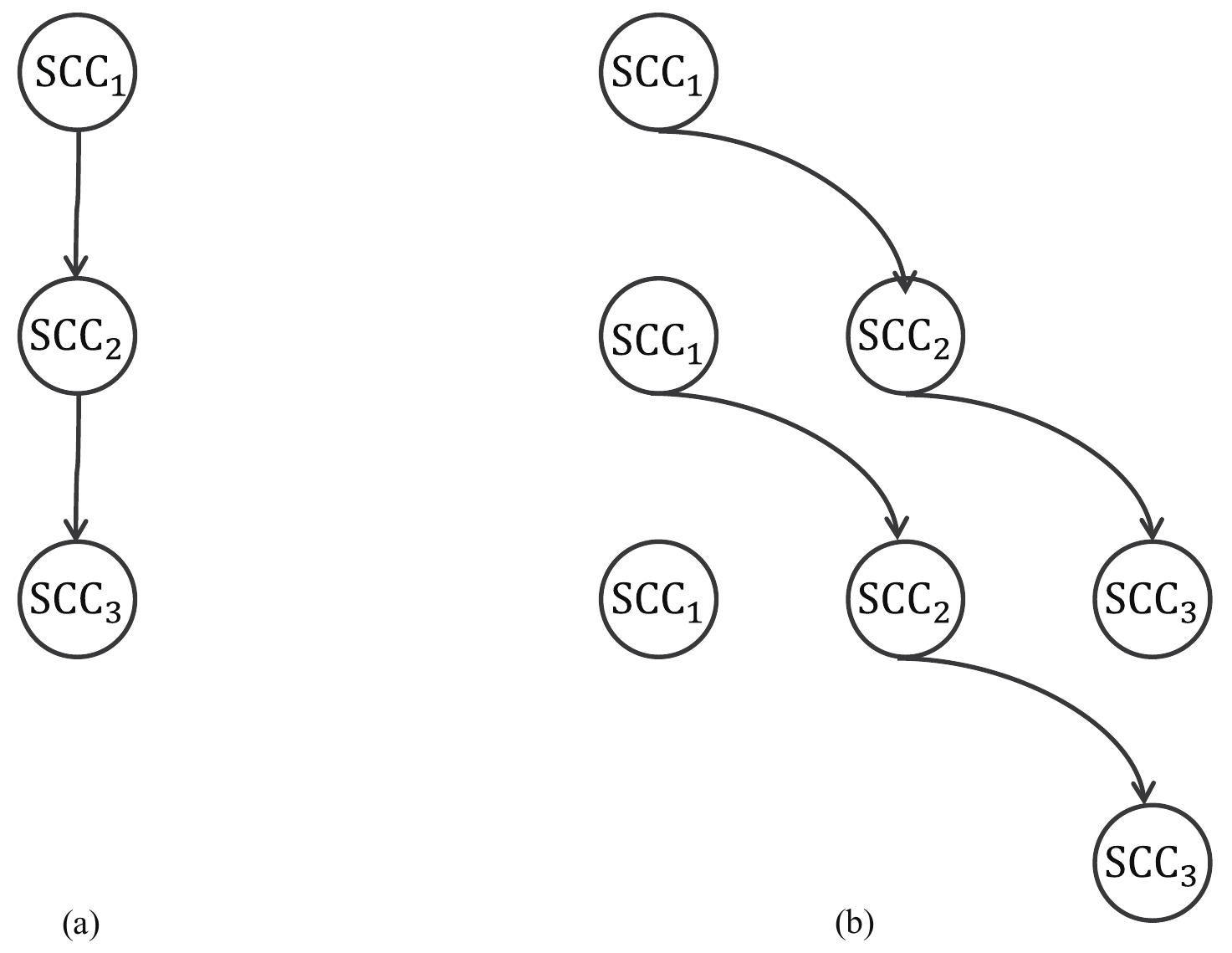}
\caption{(a) A dependence graph with SCC contracted into nodes. (b) A pipelined execution of the SCC across three threads.}
\label{fig4}
\end{figure}

Cyclic dependence graphs with at least one loop-carried dependence, and the statement will form a SCC in the dependence graph.
The most straightforward way to deal with the statement in each SCC is to place in a loop that is executed sequentially. Another way of extracting parallelism from these loops is to execute the SCC in a pipelined fashion. An example of this is
shown in Fig.\ref{fig4}. This is called $decoupled$ $software$ $pipelining$, and is described in detail in \cite{20}.

In latter section 4, we show how parallelism can sometimes be extracted from these loops using $producer-consumer$ synchronization, and optimizing  $producer-consumer$ synchronization.

\section{Optimizing Synchronization Algorithm}
There is no guarantee the order that parallel program execute on the different threads will enforce the dependence. However, by using producer/consumer synchronization, this ordering can be forced on the program execution, allowing parallelism to be extracted from loops with dependence.

In the 1980s and early 1990s, several forms of producer/consumer synchronization were implemented (e.g. full/empty synchronization, implemented in the Denelcor HEP \cite{21}). The Alliant F/X 8 \cite{8} \cite{9} implemented the $advance(r,i)$ and $await(r,i)$ synchronization instructions. In 1987, Samuel P. Midkiff discussed the compiler algorithms
for synchronization \cite{22}.  He explained $\boldsymbol{with}$, $\boldsymbol{quit}$, $\boldsymbol{test}$, $\boldsymbol{testset}$, $\boldsymbol{wait}$, and $\boldsymbol{set}$ instructions in detail.

In this section, compiler exploitation of both of these synchronization instruction, and general producer/consumer synchronization, can be discussed in terms of $\boldsymbol{send}$ and $\boldsymbol{wait}$ synchronization. The $\boldsymbol{wait}(regs, i, vars)$ waits until the value of $regs$ is $i$. The $\boldsymbol{send}(regs, i, vars)$
writes the value $i$ to $regs$, where $i$ is the loop index variable, $regs$ is the synchronization register used for dependence $\delta$, and $vars$ contains the variables involved whose dependence is being synchronized. The $\boldsymbol{send}$ and $\boldsymbol{wait}$ instructions also have a functionality equivalent to a $\boldsymbol{fence}$ instruction, which would ensure that result of all memory accesses before the $\boldsymbol{send}$ and $\boldsymbol{wait}$ are visible before the $\boldsymbol{send}$ or $\boldsymbol{wait}$ competes, and
the hardware doesn't move instructions past the synchronization operation at run time.
\subsection{Insert Synchronize Instruction Set}

Due to the dependence graph, a compiler can synchronize a program directly. In order to a deep understanding, there
is an example of using producer/comsumer synchronization, and the program is simplified as Alg.\ref{alg4}. If you observe keenly, it's easy to find out the dependence graph for the program (i.e. $\delta^f,\Delta_a = 1$; $\delta^f,\Delta_b = 2$; $\delta^f,\Delta_c = 1$).

\begin{algorithm}
\caption{A loop with cross-iteration dependence.} \label{alg4}
\begin{algorithmic}
       \FOR{$i=1$; $i<n$; $i++$ }
          \item $S1: \space a[i] \gets b[i-1]+... ;$\\
           \item $S2: \space b[i] \gets c[i-1]+... ;$\\
           \item $S3: \space c[i] \gets b[i-2]+a[i-1] ;$\\
         \ENDFOR
\end{algorithmic}
\end{algorithm}

\begin{figure}
\centering
\includegraphics[width=0.25\textwidth]{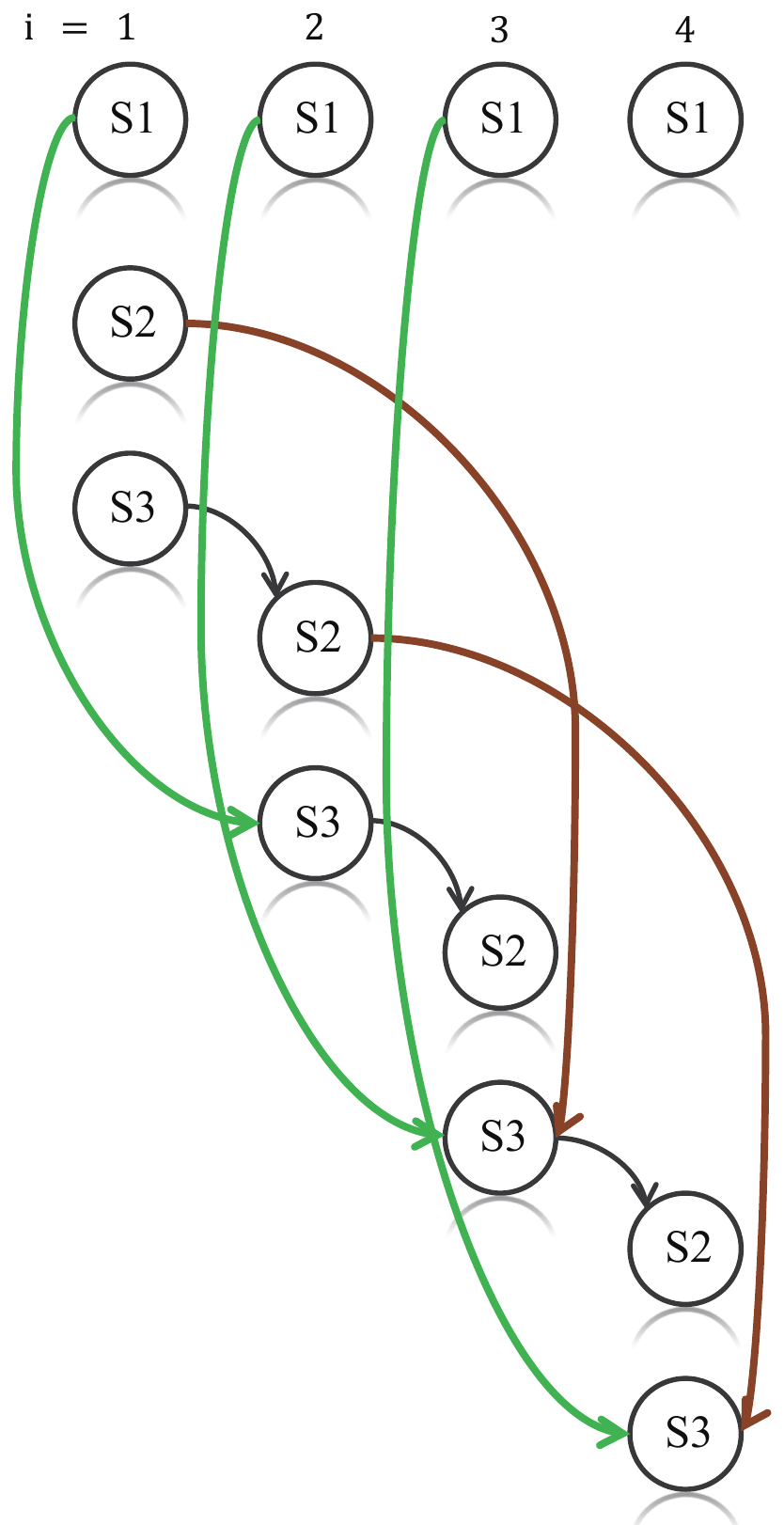}
\caption{the iteration space of the loop of Alg.\ref{alg4}.}
\label{fig6}
\end{figure}

When we know the dependence distance from the dependence graph, the iteration space of the loop of the program can be illustrated in Figure.\ref{fig6}.


The iteration space can make ensure the location of the synchronize instructions. As you see, the green dotted line denotes the $\delta^f,\Delta_a = 1$, the brown dotted line denotes the $\delta^f,\Delta_b = 2$, and  the solid line denotes $\delta^f,\Delta_c = 1$. After the source of dependence $\delta$, it inserts the instruction $send(regs_\delta, i, vars)$. Before each dependence sink, the compiler inserts the instruction $wait(regs_\delta, i-d_j, vars)$, where $d_i$ is the distance of the dependence on the $i$ loop. The loop of the program synchronized with $\boldsymbol{send}$/$\boldsymbol{wait}$
synchronization has be shown in Alg.\ref{alg5}.

\begin{algorithm}
\caption{A loop of the program synchronized with $\boldsymbol{send}$/$\boldsymbol{wait}$ synchronization.}
\label{alg5}
\begin{algorithmic}
       \FOR{$i=1$; $i<n$; $i++$ }
          \item $S1: \space a[i] \gets b[i-1]+... ;$\\
          $\boldsymbol{send}$(0, i, a);\\
          $\boldsymbol{wait}$(2, i-1, c);\\
           \item $S2: \space b[i] \gets c[i-1]+... ;$\\
          $\boldsymbol{send}$(1, i, b);\\
          $\boldsymbol{wait}$(1, i-2, b);\\
          $\boldsymbol{wait}$(0, i-1, a);\\
           \item $S3: \space c[i] \gets b[i-2]+a[i-1] ;$\\
           $\boldsymbol{send}$(2, i, c);\\
         \ENDFOR
\end{algorithmic}
\end{algorithm}

The reasons that producer/consumer synchronization instructions aren't supported in hardware anymore shows that impact that technology and economics dependent on what is a desirable architectural \cite{23}. Specialized synchronizing instructions fell out of favor because of the increased latencies required when synchronizing across the system bus between general purpose processors, and because the RISC principles of instruction set design \cite{24} favored simpler instructions from which $\boldsymbol{send}$ and $\boldsymbol{wait}$ instructions could be built, albeit at a higher run time cost. Except for questions of profitability, the compiler strategy for inserting and optimizing synchronization is indifferent to whether it is implement in software or hardware. These optimizes will be explained in detail in the remainder of this section.

\subsection{Two Approaches to Optimize Synchronization}
Sometimes a compiler may reduce the number of synchronization operations needed to synchronize the dependence in a loop.
However, all dependence must be enforced, So this optimization reduces the overhead of enforcing them by allowing a single
$\boldsymbol{send}$/$\boldsymbol{wait}$ pair to synchronize more than one dependence, or a combination of $\boldsymbol{send}$/$\boldsymbol{wait}$ instructions to synchronize additional dependence. There is a loop with dependence to be synchronized in Alg.\ref{alg6}

\begin{algorithm}
\caption{A loop with dependence to be synchronized.}
\label{alg6}
\begin{algorithmic}
       \FOR{$i=1$; $i<n$; $i++$ }
          \item $S1: \space a[i] \gets ... ;$\\
          \item $S2: \space b[i] \gets c[i-1]+... ;$\\
           \item $S3: \space c[i] \gets a[i-2] ;$\\
         \ENDFOR
\end{algorithmic}
\end{algorithm}

\begin{figure}
\centering
\includegraphics[width=0.40\textwidth]{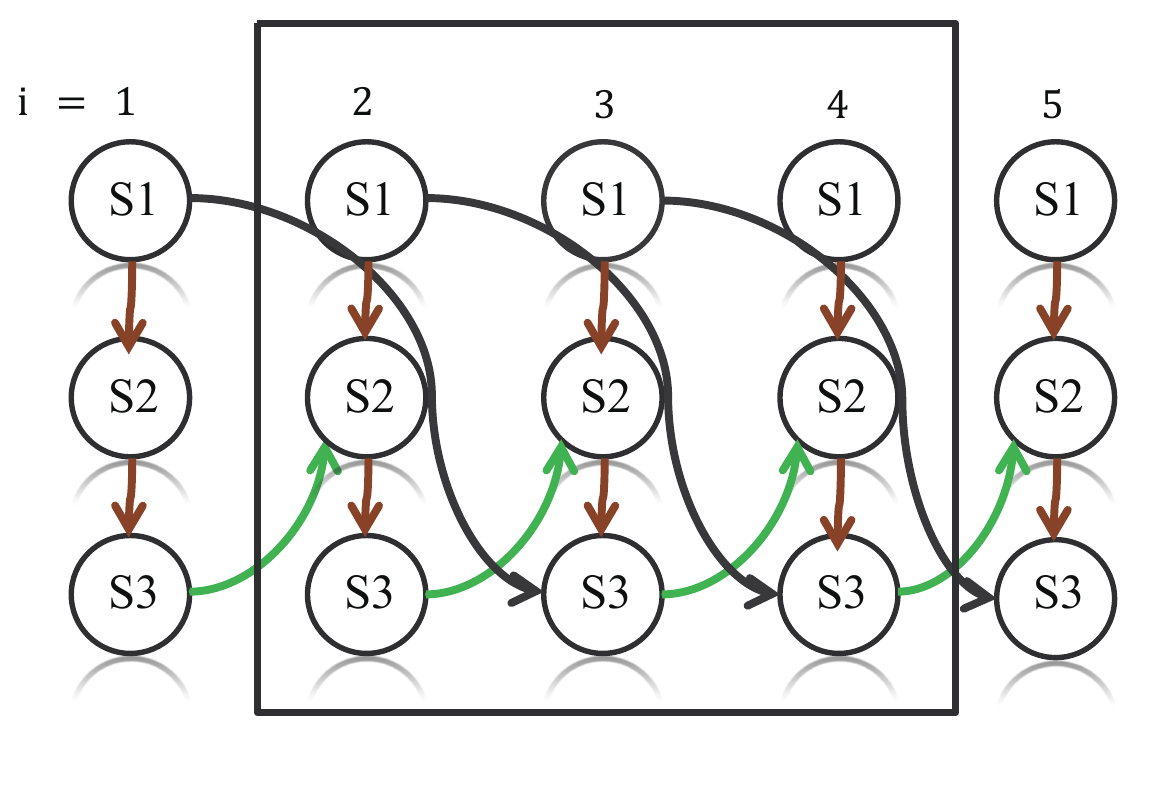}
\caption{the ISD for the loop of Alg.\ref{alg6}.}
\label{fig7}
\end{figure}

The loop with two dependence, include $\delta^f,\Delta_a = 2$ and $\delta^f,\Delta_c = 1$. The $iteration space diagram$(ISD) of Figure.\ref{fig7} shows the dependence to be enforced as the blue solid lines or the green dashed lines, and execution orders implied by the sequential execution of the program by the brown dashed lines. The section outlined with dotted box is representative of a section of the ISD that is examined by the algorithm of \cite{10} that eliminates dependence using transitive reduction.

Let $S_j(k)$ represent the instance of statement $S_j$ in iteration $i = k$. Consider the dependence with distance two from statement $S_1$ in iteration $i = 2$ to statement $S_3$ in iteration $i = 4$. There is a path $S_1(2)$ $\rightarrow$ $S_2(2)$
$\rightarrow$ $S_3(2)$ $\rightarrow$ $S_2(3)$ $\rightarrow$ $S_3(3)$ $\rightarrow$ $S_2(4)$ $\rightarrow$ $S_3(4)$ from $S_1$
 in iteration 2 to $S_3$ in iteration 4, just like the black lines in the dotted box. If the dependence from $S_3$ to $S_2$ has been synchronized, then the existence of this path of enforced orders implies that the dependence from $S_1(2)$ to $S_3(4)$ is also enforced. Due to the distances are constant, the iteration space can be covered by shifting the region in the dashed lines, So every instance of the dependence within the iteration space is synchronized. Samuel P. Midkiff had already shown that  perform a transitive reduction on the ISD \cite{10}. It's possible for multiple dependence to work together to eliminate another dependence. The transitive reduction is performed on the ISD, which needs to only contain a subset of the total iteration space (i.e. the case as shown by the dotted box in Figure.\ref{fig7}). For each loop in the loop nest over which the synchronization elimination is taking place, the number of iterations needed in the ISD for the loop
is equal to the least product of the unique prime factors of the dependence distance, plus one.

Another synchronization elimination approach \cite{25} is based on pattern matching and works even if the dependence distance are not constant. The matched patterns identify dependence whose lexical relationship and distance are such that synchronizing one dependence will synchronize the order by forming a path as shown in Figure \ref{fig7} (i.e. the black lines in the dotted box). In the program of Alg.\ref{alg6}, let the forward dependence with a distance of two that is to be eliminated be $\delta_e$, and the backward dependence of distance one be $\delta_1$ that is used to be eliminated the other dependence be $\delta_r$. There is one pattern as follows:

\begin{itemize}
\item[i]
A path from the source of $\delta_e$ to the source of some $\delta_r$.
\item[ii]
The sink of $\delta_r$ reaches the sink of $\delta_e$.
\item[iii]
$\delta_r$ is lexically backward (i.e. the sink precedes the source in the program flow).
\item[iv]
The absolute value of the distance of $\delta_r$ is one.
\item[v]
The signs of the distances of $\delta_e$ and $\delta_r$ are the same, then $\delta_e$ can be eliminated.
\end{itemize}

The conditions of i and ii establish the proper flow of $\delta_e$ and $\delta_r$, the iii recognizes that $\delta_r$
can be repeatedly executed to reach all iterations that are multiple of the distance away from the source. The iv and v show that because the absolute value of the distance is one and the signs of the two distances are equal the traversal enabled by the iii will reach the source of $\delta_e$.

\section{Conclusion}
We have studied the way of the $\boldsymbol{send}$ and $\boldsymbol{wait}$ instructions to synchronize loops. We have given general strategies for treating branches within a loop being synchronized, and present
two approaches to reduce and optimize the number of producer/consumer synchronization instructions in the shared-memory multiprocessors machine.

In general, when synchronized the version of parallel program, there are four steps need to be enforced.
First, a dependence graph is illustrated with respect to the program. Second, depending on the structure of the dependence graph and the relative costs of the different synchronization methods on a target machine, Picking a synchronization method to synchronize the loop. Third, synchronize instructions are inserted, and
it makes sure that the cross-iteration dependence can be enforced. Finally, eliminating partial dependence  and optimizing the synchronize instructions.

Auto-parallelizing compiler can perform all of these steps automatically, which relieve programmers from the tedious and error-prone manual parallel process.


\begin{thebibliography}{1}

\bibitem[1]{1}
J. M. Tendler, J. S. Dodson, J. S. Fields, Jr., H. Le, and B. Sinharoy. POWER4 system microarchitecture. IBM Journal of Research and Development 46 (1): 5-26. 2002.
%

\bibitem[2]{2}
Swinburne, Richard. Intel Core i7 - Nehalem Architecture Dive. 5 - Architecture Enhancements. 2008.
%

%
\bibitem[3]{3}
D. F. Bacon, S. L. Graham, and O. J. Sharp. Compiler transformations for high-performance computing. ACM Comput. Surv., 26:345-420, December 1994.
%

%
\bibitem[4]{4}
L Dagum, R Menon. OpenMP: an industry standard API for shared-memory programming. Computational Science \& Engineering, IEEE. Jan-Mar 1998.
%

%
\bibitem[5]{5}
D. J. Kuck, R. H. Kuhn, D. A. Padua, B. Leasure, and M.Wolfe. Dependence graphs and
compiler optimizations. In ACM Conference on the Principles of Programming Languages, pages
207-218, 1981.
%

%
\bibitem[6]{6}
David J. Kuck. A Survey of Parallel Machine Organization and Programming. ACM Computing Surveys - CSUR, vol. 9, no. 1, pp. 29-59, 1977
%

%
\bibitem[7]{7}
Randy Allen, Ken Kennedy. Optimizing Compilers for Modern Architectures: A Dependence-based Approach. Morgan Kaufmann. ISBN 1-55860-286-0. 2001.
%


%
\bibitem[8]{8}
J. Test, M. Myszewski, and R. Swift. The alliant fx/series: A language driven architecture for
parallel processing of dusty deck fortran. In J. de Bakker,A.Nijman, and P.Treleaven, editors,
PARLE Parallel Architectures and Languages Europe, volume 258 of Lecture Notes in Computer
Science, pages 345¨C-356. Springer Berlin / Heidelberg, 1987.
%

%
\bibitem[9]{9}
W. A. Abu-Sufah and A. D. Malony. Vector processing on the Alliant FX/8 multiprocessor.
In Proceedings of the International Conference on Parallel Processing, pages 559-566, 1986.
%

%
\bibitem[10]{10}
S.P.Midkiff and D.A.Padua. Compiler generated synchronization for do loops. In Proceedings
of the International Conference on Parallel Programming (ICPP), pages 544-551, 1986.
%
\bibitem[11]{11}
M. Bach, M. Charney, R. Cohn, T. Devor, E. Demikovsky, K. Hazelwood, A. Jaleel, C.-K.
Luk, G. Lyons, H. Patil, and A.Tal. Analyzing parallel programs with pin. IEEE Computer,
43(3):34-41, March 2010.
%
\bibitem[12]{12}
Alfred V.Aho, Monica S.Lam, Ravi Sethi, and Jeffrey D.Ullman. Compilers Pricinples, Techniques, \& Tools 2nd
ed, ISBN 978-7-111-32674-8, 2006.

\bibitem[13]{13}
R. Cytron, J. Ferrante, B.K. Rosen, M.N.Wegman, and F. K.Zadeck. An efficient method of
computing static single assignment form. In ACM Conference on the Principals of Programming
Languages, pages 25-35, 1989.

\bibitem[14]{14}
John L. Hennessy, David A. Patterson. Computer Architecture: A Quantitative Approach 5th ed, ISBN 978-7-111-36458-0, 2012.

\bibitem[15]{15}
D. A. Padua, D. J. Wolfe. Advanced compiler optimizations for supercomputers, Commum. ACM, vol. 29, pp. 1184-1201, Dec. 1986.

\bibitem[16]{16}
M.J.Wolfe. Optimizing supercompilers for supercomputers, Ph.D. dissertation, Univ. Illinois, Urbana-Champaign, DCS Rep. UIUCDCS-R-82-1105, Oct. 1982.


\bibitem[17]{17}
Samuel P. Midkiff. Automatic Parallelization: An Overview of Fundamental Compiler Techniques, Morgan \& Claypool, Feb, 2012

\bibitem[18]{18}
Thomas H. Cormen, Charles E. Leiserson, Ronald L. Rivest, Clifford Stein. Introduction to Algorithms 3ed, The MIT Press, pp. 612-615, ISBN 978-0-262-53305-8, 2009

\bibitem[19]{19}
M. Wolfe. High performance compilers for parallel computing. Addison-Wesley Publishing Company, 1996.

\bibitem[20]{20}
E. Raman, G. Ottoni, A. Raman, M. J. Bridges, and D. I. August. Parallel-stage decoupled software pipelining. In Proceedings of the 6th annual IEEE/ACM International Symposium on Code generation and optimization, CGO ¡¯08, pages 114-123, New York, NY, USA, 2008.

\bibitem[21]{21}
J. S. Kowalik. Parallel MIMD computation. The HEP supercomputer and its applications. MIT Press, 1985.


\bibitem[22]{22}
Samuel P. Midkiff: Compiler Algorithms for Synchronization. IEEE Transctions On Computers, Vol. C-36, No.12, 1987


\bibitem[23]{23}
E. B. III and K. Warren. The 1991 MPCI yearly report: The attack of the killer micros. Technical report,Lawrence LivermoreNational Laboratory, 1991.

\bibitem[24]{24}
D. A. Patterson, C. H. Sequin. RISC I: a reduced instruction set VLSI computer. In 25
years of the International Symposia on Computer Architecture, ACM., pp. 216¨C230,
New York, NY, USA, 1998.

\bibitem[25]{25}
Z. Li and W. A. Abu-Sufah. A technique for reducing synchronization overhead in large
scale multiprocessors. In Proceedings of the International Symposia on Computer Architecture,
pages 284¨C291, 1985.
\end{thebibliography}
\end{document}